# Reconstruction of missing information in diffraction patterns and holograms by iterative phase retrieval

Tatiana Latychevskaia

University of Zurich, Winterthurerstrasse 190, 8057 Zurich, Switzerland

tatiana@physik.uzh.ch

**ABSTRACT**

It is demonstrated that an object distribution can be successfully retrieved from its diffraction pattern or hologram, even if some of the measured intensity samples are missing. The maximum allowable number of missing values depends on the linear oversampling ratio σ, where the higher the value of $\sigma$, the more intensity samples can be missing. For a real-valued object, the ratio of missing pixels to the total number of pixels should not exceed $(1 - 2/\sigma^2)$ or $(1 - 1/\sigma^2)$ in the acquired diffraction pattern or hologram, respectively. For example, even 5% of the measured intensity values at an oversampling ratio of $\sigma = 8$ are sufficient to simultaneously retrieve the object distribution and the missing intensity values. It is important that the missing intensity values should not be concentrated in the centre, but should be randomly distributed over the acquired diffraction pattern.

## Contents

# 1. Introduction

Technological developments in information theory have always aimed to optimise information transfer, minimising the number of measurements that allow for capture of the complete signal distribution. For example, the Nyquist-Shannon-Kotelnikov theorem [1, 2] gives the minimal sampling rate at which sample measurements completely determine the signal. Similar observations have been made in optics, namely in coherent diffraction imaging (CDI) [3] and holography [4, 5]. In CDI and holography, the intensity distribution of the scattered wave is acquired by a distant detector (thus producing a diffraction pattern or a hologram, respectively) and the object distribution is then reconstructed from the acquired intensity distribution (the diffraction pattern or the hologram, respectively). In practice, detectors may have some faulty pixels that deliver incorrect values. For example, this may include pixels that deliver a zero intensity value ("dead" pixels [6]) or those which deliver an extremely high intensity value (saturated or "bright" pixels). In X-ray [7-10] or electron [11] diffraction experiments, the signal in the central part of the diffraction pattern is often missing due to a beamstop, or is overexposed due to a direct beam or a hole in the detector. Although the correct information in these pixels is missing, it can be recovered during the iterative phase retrieval reconstruction procedure. It has been noted that although the acquired intensity distribution may be incomplete, or in other words have missing intensity values, a good quality reconstruction of the object can still be achieved [6, 12]. This study attempts to answer the question of how much information can be missing from the acquired diffraction pattern or hologram so that the imaged object can be still reconstructed without error by using conventional iterative phase retrieval algorithms.

## 2."Missing" intensity values in coherent diffraction imaging

### 2.1. Oversampling condition and "missing" intensity values

The principles of CDI and "oversampling" are well explained in the existing literature [13], and here we only mention a few points related to the present study. The intensity of the wavefront scattered from the object $f(\vec{r})$ measured in the far field provides the values of the magnitude of the Fourier transform of the object distribution:

$$|F(\vec{u})| = \left| \sum_{\vec{r}=0}^{N-1} f(\vec{r}) \exp(-i\vec{u} \cdot \vec{r}) \right|, \quad (1)$$

which is a set of equations where $f(\vec{r})$ are the unknowns. We consider a 2D diffraction pattern $I(\vec{u}) = |F(\vec{u})|^2$ sampled with $N \times N$ pixels. The total object area is also sampled with $N \times N$ pixels, and the object distribution is sampled with $N_0 \times N_0$ pixels. This gives for a complex-valued object, there are $N^2$ equations and $2N_0^2$ unknowns; for a real-valued object, there are $N^2/2$ equations and $N_0^2$ unknowns [13]. Thus, the system of equations can in principle have a solution if the number of equations exceeds the number of unknowns. For real-valued objects, this condition is:

$$\frac{N^2}{2} > N_0^2, \quad (2)$$

which can be re-written by introducing the *linear* oversampling ratio

$$\sigma = \frac{N}{N_0} \quad (3)$$

as

$$\sigma > \sqrt{2}, \quad (4)$$

which is the oversampling condition. For the 1D case, the linear oversampling ratio should satisfy $\sigma > 2$, and for the 3D case the linear oversampling ratio should satisfy $\sigma > 2^{1/3}$ [13]. The linear oversampling ratio means that the oversampling condition should be fulfilled in each dimension. When the oversampling condition is not fulfilled in any of the dimensions, the object distribution cannot be reconstructed, as previously demonstrated in [12].

A CDI experiment is prepared in such a way that the oversampling condition is fulfilled, and a certain oversampling ratio is achieved [12, 13]. The extent of the reconstructed area $S_0 \times S_0$ is provided from the Fourier transform: $S_0 = \frac{\lambda z}{\Delta}$, where the pixel size in the detector plane $\Delta$ is given by $\Delta = \dfrac{S}{N}$, $S \times S$ is the detector size, $N \times N$ is the number of pixels, $\lambda$ is the wavelength, and $z$ is the distance between the object and the detector. When the diffraction pattern is measured in the $k$-domain, the extent of the reconstructed area is given by $S_0 = \dfrac{2\pi}{\Delta_k}$, where $\Delta_k$ is the pixel size in the $k$-domain. The object extent $O$ is approximately known. The linear oversampling ratio is then given by $\sigma = \dfrac{S_0}{O}$.

The object distribution can be reconstructed from its diffraction pattern provided the latter is sampled at twice the Nyquist frequency [13, 14]. If some of the intensity measurements in the measured diffraction pattern are missing, the number of missing pixels can be characterised by

$$f = \frac{N_{missing}}{N_{total}}, \tag{5}$$

where $f$ is the ratio of the missing pixels to the total number of pixels in the diffraction pattern. For a 2D signal, $N_{total} = N^2$. This gives the number of measured pixels as $(1-f)N^2$, and Eq. 2 can be re-written as:

$$(1-f)\frac{N^2}{2} > N_0^2. \tag{6}$$

Using the definition of the oversampling ratio in Eq. 3, we obtain the following condition for the missing pixel ratio:

$$f < 1 - \frac{2}{\sigma^2}. \tag{7}$$

From Eq. 7, for example, for $\sigma = 4$, we obtain $f < 0.875$ and for $\sigma = 8$ we obtain $f < 0.969$. This means that for a diffraction pattern with a linear oversampling ratio $\sigma = 8$, even 4% of all intensity measurements are in principle sufficient to reconstruct the object distribution and simultaneously recover the intensity values at the missing pixels.

## 2.2. Simulated examples

The diffraction pattern of an amplitude object was simulated as described in Appendix A. Next, at randomly distributed coordinates, the intensity values in the simulated diffraction pattern were set to zero. Such pixels are considered to be missing pixels. The ratio of the missing pixels to the total number of pixels in the diffraction pattern is characterised by the factor given in Eq. 5. An example of a diffraction pattern with 50% of the pixels missing ($f = 0.5$) is shown in Fig. 1.

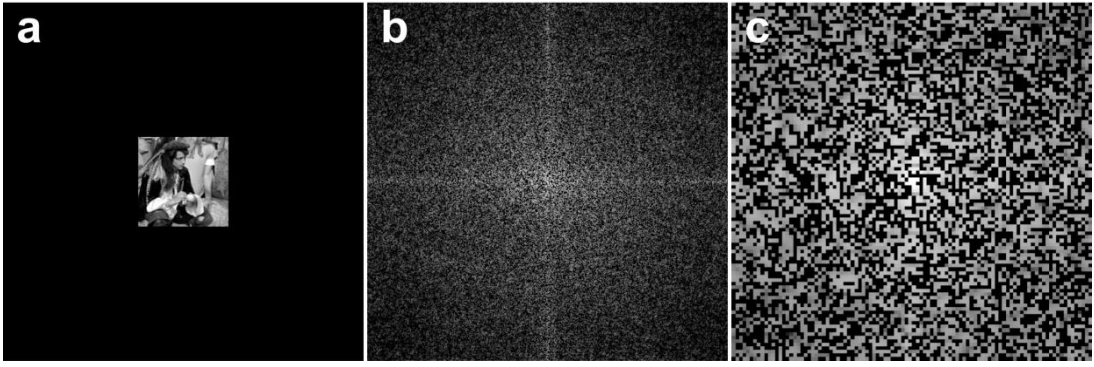

Fig. 1. Randomly distributed pixels missing from a diffraction pattern. (a) The total distribution in the object plane, sampled with 512 × 512 pixels. The central part of the object distributions is the "man" object, sampled with 128 × 128 pixels. (b) The simulated diffraction pattern sampled with 512 × 512 pixels, at an oversampling ratio of 4. Here, 50% of the pixels are missing, and thus $f = 0.5$. (c) The magnified central 100 × 100 pixels area of the diffraction pattern shown in (b). The intensity distributions in (b) and (c) are shown on a logarithmic scale.

The protocol for the iterative reconstruction of diffraction patterns, with recovery of missing pixels, is provided in Appendix A. The mismatch between the reconstructed and the original object distributions ("man" image, non zero-padded) is calculated as:

$$E = \frac{1}{N_0} \sqrt{\sum_{x,y=1}^{N_0} |o(x,y) - o_0(x,y)|^2}, \tag{8}$$

where $o(x, y)$ is the reconstructed object distribution, $o_0(x, y)$ is the original object distribution, sampled with $N_0 \times N_0$, and $x, y = 1...N_0$ are the pixel coordinates. In the examples shown below, the object distribution is sampled with $N_0 \times N_0 = 128 \times 128$ pixels, and is zero-padded to $N \times N = 512 \times 512$ or $N \times N = 1024 \times 1024$ to achieve linear oversampling ratios of 4 and 8, respectively.

The numerical simulations shown below are carried out for noise-free diffraction patterns. The effect of noise on the quality of the reconstructions obtained from diffraction patterns was recently studied in [12]. The effect of the noise is not considered in the present study for two reasons. (1) In order to provide a good study of the noise problem in addition to the missing pixel problem, different levels of noise need to be added to each diffraction pattern with missing pixels. This would not only be time consuming, but it would be also impossible to present such a large amount of calculations in one paper. (2) Experimentally, there are ways to reduce noise by simply using a longer acquisition time or a more intense incident beam.

### 2.2.1. Randomly distributed missing pixels

The object distribution ("man") was sampled with 128 × 128 pixels and zero-padded to 512 × 512 pixels. The resulting simulated diffraction pattern is thus oversampled with a linear oversampling ratio of 4. The reconstructions are shown in Fig. 2 in pairs: on the left is the reconstruction obtained from the complex-valued (in other words, the phase distribution is available) far-field distribution by taking the inverse Fourier transform (FT), and on the right is the reconstruction obtained from the far-field diffraction pattern by applying the iterative phase retrieval routine. From the results shown in Fig. 2, it is evident that the quality of the reconstructed image worsens when the ratio of missing pixels to the total number of pixels $f$ increases. Starting from $f = 0.6$, which means that 60% of the diffraction pattern's pixels are missing, the reconstructed distributions do not visually resemble the original distribution. Moreover, the dependency of the error as a function of iteration exhibits a stagnation (Fig. 2u), and a larger number of iterations will probably not lead to a better reconstruction. The errors in the obtained reconstructions are summarised in Table 1.

In Fig. 2 and Table 1 we can observe certain inconsistencies between the quality of the reconstruction (Fig. 2) and the corresponding error (Table 1). The reason for this is as follows. An inverse Fourier transform is performed on complex-valued far-field distributions with missing pixels. These missing pixels reduce the total amplitude of the far-field distribution, and as a result, the amplitude of the reconstructed object distribution is also reduced in accordance with Parseval's theorem. Thus, even though the reconstruction looks correct to the naked eye, the amplitude of the reconstruction may be several times lower than the original distribution, which in turn results in a large error according to Eq. 8. The reconstructions obtained from the iterative phase retrieval process do not suffer from this problem, since the signal in the missing pixels is almost completely recovered, and therefore the original values of the amplitude in both the diffraction pattern and the object distribution are reconstructed. To demonstrate this phenomenon, the amplitude values of the reconstructed object distributions are indicated in the corresponding figures.

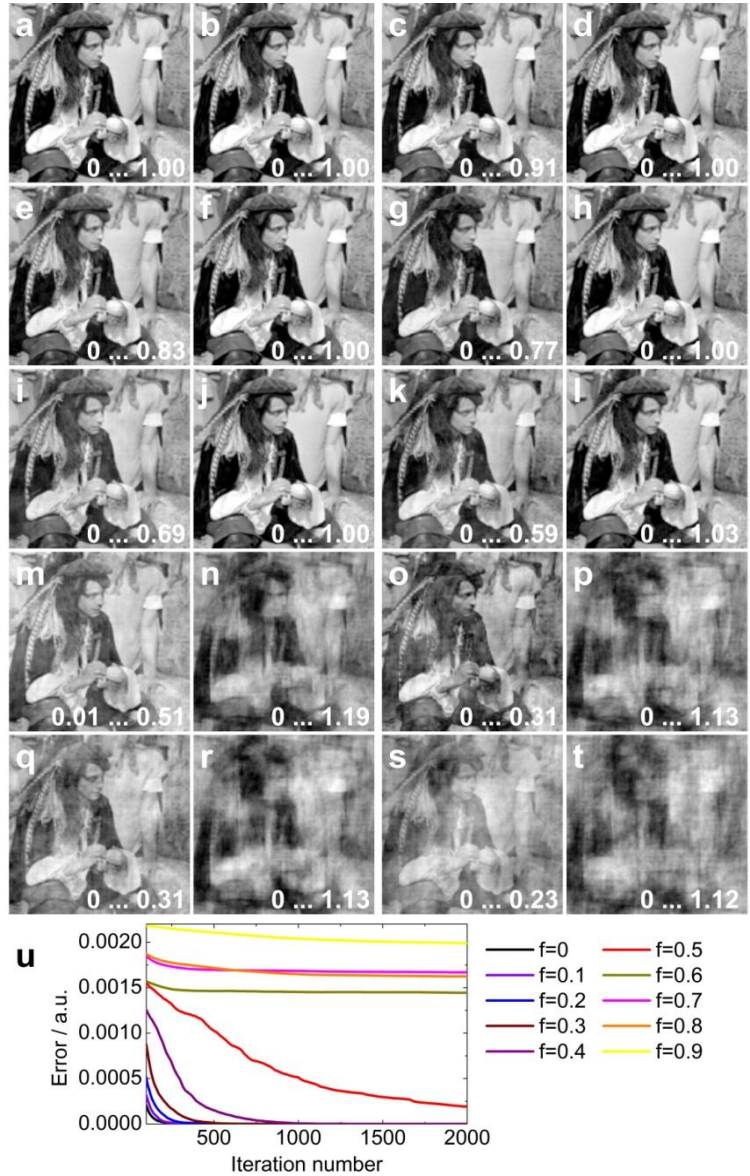

Fig. 2. Missing pixels in the diffraction pattern of a real-valued object, where the linear oversampling ratio is 4. The central part of the reconstructed object distributions, 128 × 128 pixels, is shown. The total reconstructed area is sampled with 512 × 512 pixels. The reconstructions are shown in pairs. Left: reconstruction obtained from the complex-valued far-field distribution by taking an inverse Fourier transform. Right: reconstruction obtained from the far-field diffraction pattern by applying the iterative phase retrieval routine. The numbers in the lower right-hand corners indicate the amplitude values of the reconstructed object distributions in a.u. The ratio of missing pixels to the total number of pixels ($f$) is (a) – (b) $f = 0$, (c) – (d) $f = 0.1$, (e) – (f) $f = 0.2$, (g) – (h) $f = 0.3$, (i) – (j) $f = 0.4$, (k) – (l) $f = 0.5$, (m) – (n) $f = 0.6$, (o) –

(p) $f = 0.7$, (q) – (r) $f = 0.8$, (s) – (t) $f = 0.9$. (u) The error as a function of the iteration number for different $f$, calculated using Eq. 8.

|            | $f=0$     | $f=0.1$  | $f=0.2$  | $f=0.3$  | $f=0.4$  |
|------------|-----------|----------|----------|----------|----------|
| Inverse FT | 2.48E-11  | 4.59E-4  | 8.24E-4  | 1.28E-3  | 1.41E-3  |
| Iterative  | 2.90E-10  | 4.91E-10 | 7.73E-10 | 1.88E-9  | 1.53E-8  |
|            | $f=0.5$   | $f=0.6$  | $f=0.7$  | $f=0.8$  | $f=0.9$  |
| Inverse FT | 1.96E-3   | 1.98E-3  | 3.06E-3  | 2.90E-3  | 3.23E-3  |
| Iterative  | 1.01E-4   | 9.38E-4  | 1.12E-3  | 1.11E-3  | 1.41E-3  |

Table 1. Error in the reconstructed object distributions calculated using Eq. 8.

The effect of the oversampling ratio is shown in Fig. 3. Here, the object distribution ("man") was sampled with 128 × 128 pixels and zero-padded to 1024 × 1024 pixels. Thus, the resulting simulated diffraction pattern was oversampled, with a linear oversampling ratio of 8. The reconstructions are shown in Fig. 3 in pairs: on the left, a reconstruction is shown that is obtained from the complex-valued far-field distribution by taking the inverse FT, and on the right is a reconstruction obtained from the far-field diffraction pattern by applying an iterative phase retrieval routine. The results shown in Fig. 3 demonstrate that a higher oversampling ratio leads to a better quality of the reconstructed images, even at relatively high value of $f$. It is only at $f = 0.8$ and $f = 0.9$ (meaning that 80% and 90% of the diffraction pattern's pixels are missing, respectively) that the reconstructed distributions do not visually resemble the original distribution. Moreover, at $f = 0.8$ and $f = 0.9$, the dependency of the error as a function of the iteration number exhibits a stagnation (Fig. 3u), and a larger number of iterations are not likely to lead to a better reconstruction. The errors in the obtained reconstructions are summarised in Table 2. In Fig. 3 and Table 2, we can observe some inconsistencies between the quality of the reconstruction (Fig. 3) and the corresponding error (Table 2). The reason for this is as discussed above for Fig. 2 and Table 1.

Comparing the results shown in Fig. 2 and 3, we conclude that a higher oversampling ratio allows for better recovery of missing information (pixels), which is in agreement with Eq. 7.

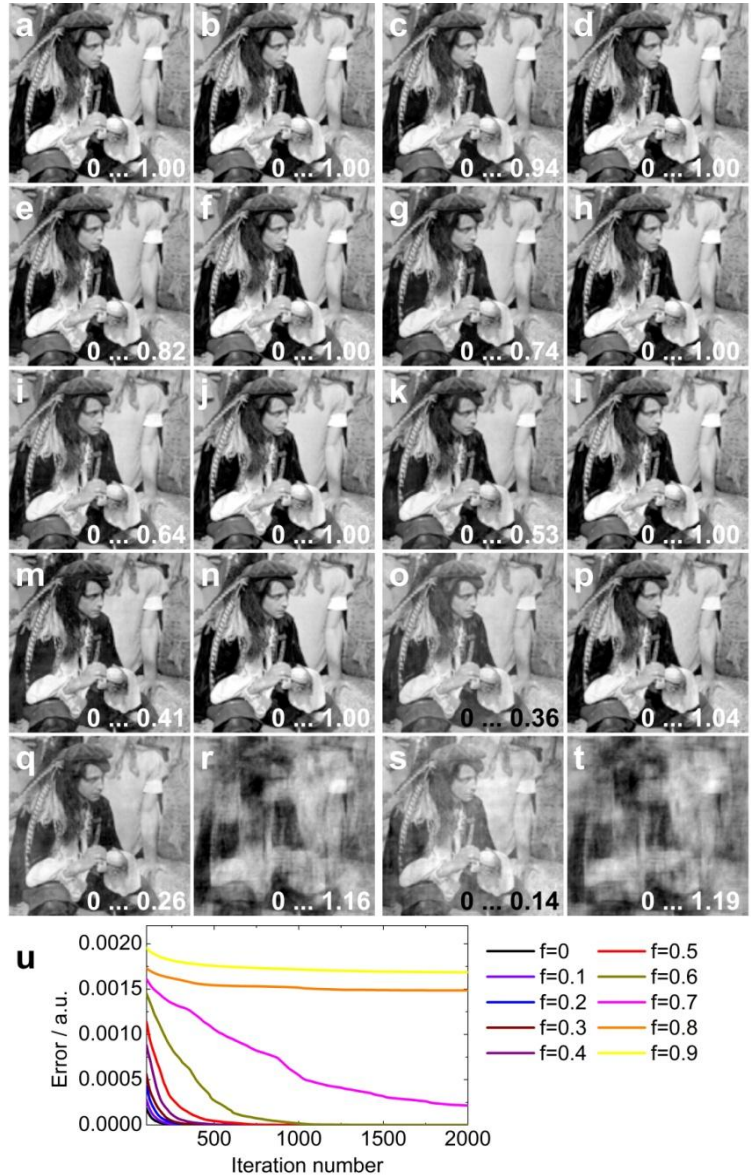

Fig. 3. Missing pixels in the diffraction pattern, where the linear oversampling ratio is 8. The central part of the reconstructed object distributions, 128 × 128 pixels, is shown. The total reconstructed object area is sampled with 1024 × 1024 pixel. The reconstructions are shown in pairs. Left: reconstruction obtained from the complex-valued far-field distribution by taking an inverse Fourier transform. Right: reconstruction obtained from the far-field diffraction pattern by applying the iterative phase retrieval routine. The numbers in the right bottom corners indicate the amplitude values of the reconstructed object distributions in a.u. The ratio of missing pixels to the total number of pixels ($f$) is (a) – (b) $f = 0$, (c) – (d) $f = 0.1$, (e) – (f) $f = 0.2$, (g) – (h) $f = 0.3$, (i) – (j) $f = 0.4$, (k) – (l) $f = 0.5$, (m) – (n) $f = 0.6$, (o) – (p) $f = 0.7$, (q)

– (r) $f = 0.8$, (s) – (t) $f = 0.9$. (u) The error as a function of the iteration number for different $f$ calculated using Eq. 8.

|            | $f=0$     | $f=0.1$   | $f=0.2$ | $f=0.3$ | $f=0.4$ |
|------------|-----------|-----------|---------|---------|---------|
| Inverse FT | 6.59E-10  | 7.48E-10  | 1.97E-9 | 3.45E-9 | 4.48E-9 |
| Iterative  | 1.67E-9   | 1.39E-9   | 3.16E-9 | 4.97E-9 | 6.44E-9 |
|            | $f=0.5$   | $f=0.6$   | $f=0.7$ | $f=0.8$ | $f=0.9$ |
| Inverse FT | 2.02E-3   | 2.59E-3   | 2.63E-3 | 3.07E-3 | 3.49E-3 |
| Iterative  | 1.52E-8   | 8.28E-8   | 1.16E-4 | 9.97E-3 | 1.11E-3 |

Table 2. Error in the reconstructed object distributions calculated using Eq. 8.

### 2.2.2 Symmetrization of diffraction pattern

In the case of a real-valued object, the corresponding diffraction pattern is centro-symmetric, and this property can be employed to reduce the number of missing pixels. A missing intensity value can be set to the value of the corresponding centro-symmetric pixel, provided that the latter has some value and is not also missing. This procedure is performed for all missing intensity values, and is referred to here as symmetrisation. This procedure reduces the overall number of missing pixels and in turn improves the quality of the reconstruction, as illustrated in Figs. 4 and 5. Figures 4 and 5 present reconstructions of the diffraction patterns shown in Fig. 2 and 3, respectively, where the number of missing pixels was reduced via the symmetrisation procedure. The object reconstructions obtained by an inverse Fourier transform of the complex-valued far-field distributions, as shown in Fig. 4 and 5, exhibit poor quality; this is because they were not symmetrised, and therefore the number of missing pixels was not reduced.

By comparing the results shown in Fig. 2 and 4, we see that the symmetrised diffraction patterns are recovered for almost all $f$ ratios, except when 90% of the pixels are missing ($f = 0.9$). The error as a function of the iteration number decreases much faster for all symmetrised diffraction patterns than for non-symmetrised diffraction patterns (compare Figs. 2u and 4u), except in the case of $f = 0.9$, where the error stagnates.

By comparing the results in Fig. 3 and 5, we see that the symmetrised diffraction patterns are successfully recovered for all $f$ ratios, even when 90% of the pixels are missing ($f = 0.9$). The error as a function of the iteration number also decreases much faster for symmetrised diffraction patterns. It also worth noting that in Fig. 5, the reconstructions obtained via the iterative phase retrieval of diffraction patterns are visually of a better quality than those obtained by inverse FT of the corresponding complex-valued far-field distributions.

The results shown in Fig. 3 and 5 also confirm the following estimations given in Eq. 6 and discussed above. For $\sigma = 4$, the object distribution can be reconstructed from its diffraction pattern if $f < 0.875$. For $\sigma = 8$, the object distribution can be reconstructed from its diffraction pattern if $f < 0.97$. However, it should be noted that successful reconstruction at these high $f$ values becomes possible only after the diffraction pattern is symmetrised.

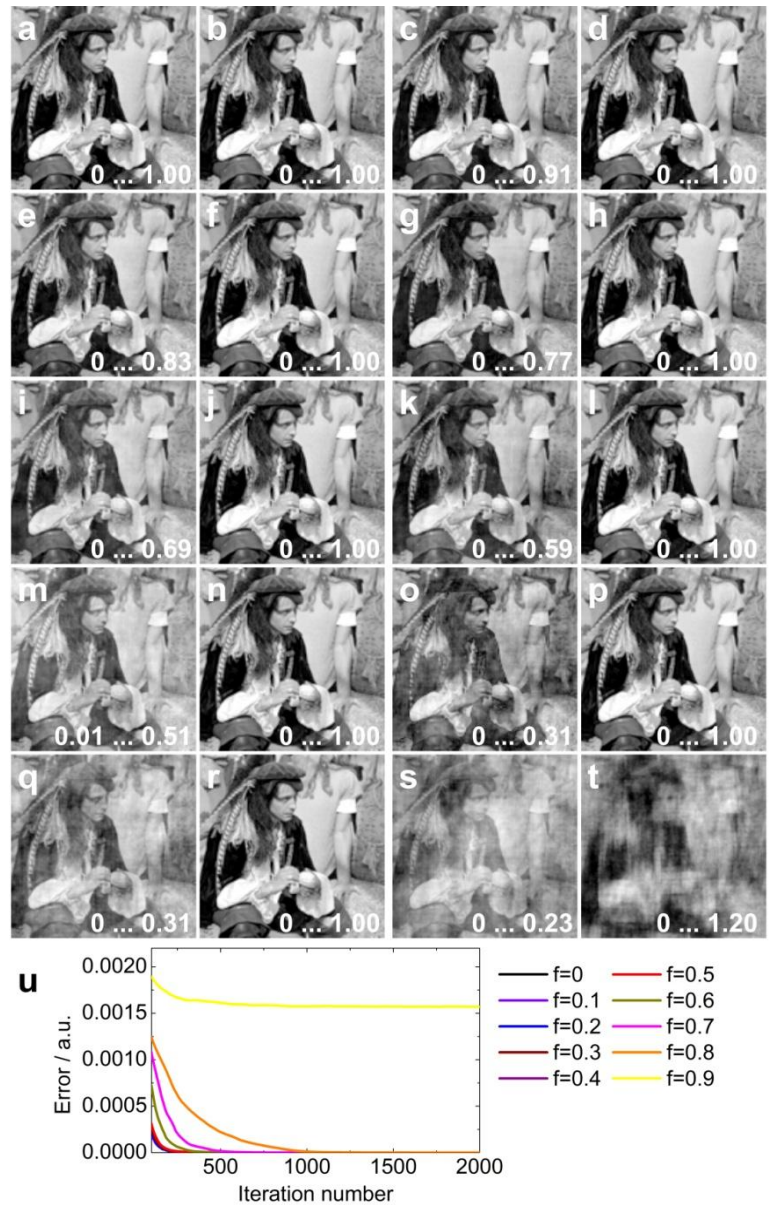

Fig. 4. Missing pixels from coherent diffraction imaging, where diffraction patterns are symmetrised before reconstruction. The central part of the reconstructed object distributions (128 × 128 pixels) is shown. The oversampling ratio is 4, and the total reconstructed object area is sampled with 512 × 512 pixels. The reconstructions are shown in pairs. Left: reconstruction obtained from the complex-valued far-field distribution by taking the inverse Fourier transform. Right: the reconstruction obtained

from the far-field diffraction pattern by applying the iterative phase retrieval routine. The numbers in the right bottom corners indicate the amplitude values of the reconstructed object distributions in a.u. The ratio of missing pixels to the total number of pixels ($f$) is (a) – (b) $f=0$, (c) – (d) $f=0.1$, (e) – (f) $f=0.2$, (g) – (h) $f=0.3$, (i) – (j) $f=0.4$, (k) – (l) $f=0.5$, (m) – (n) $f=0.6$, (o) – (p) $f=0.7$, (q) – (r) $f=0.8$, (s) – (t) $f=0.9$. (u) The error as a function of iteration number for different values of $f$, calculated using Eq. 8.

|  | $f=0$ | $f=0.1$ | $f=0.2$ | $f=0.3$ | $f=0.4$ |
|---|---|---|---|---|---|
| Inverse FT | 2.48E-11 | 4.59E-4 | 8.24E-4 | 1.28E-3 | 1.41E-3 |
| Iterative | 2.90E-10 | 3.02E-10 | 3.10E-10 | 4.38E-10 | 3.75E-10 |
|  | $f=0.5$ | $f=0.6$ | $f=0.7$ | $f=0.8$ | $f=0.9$ |
| Inverse FT | 1.96E-3 | 1.98E-3 | 3.06E-3 | 2.90E-3 | 3.23E-3 |
| Iterative | 5.05E-10 | 5.24E-10 | 2.99E-9 | 2.70E-8 | 1.06E-3 |

Table 3. Error in the reconstructed object distributions calculated using Eq. 8.

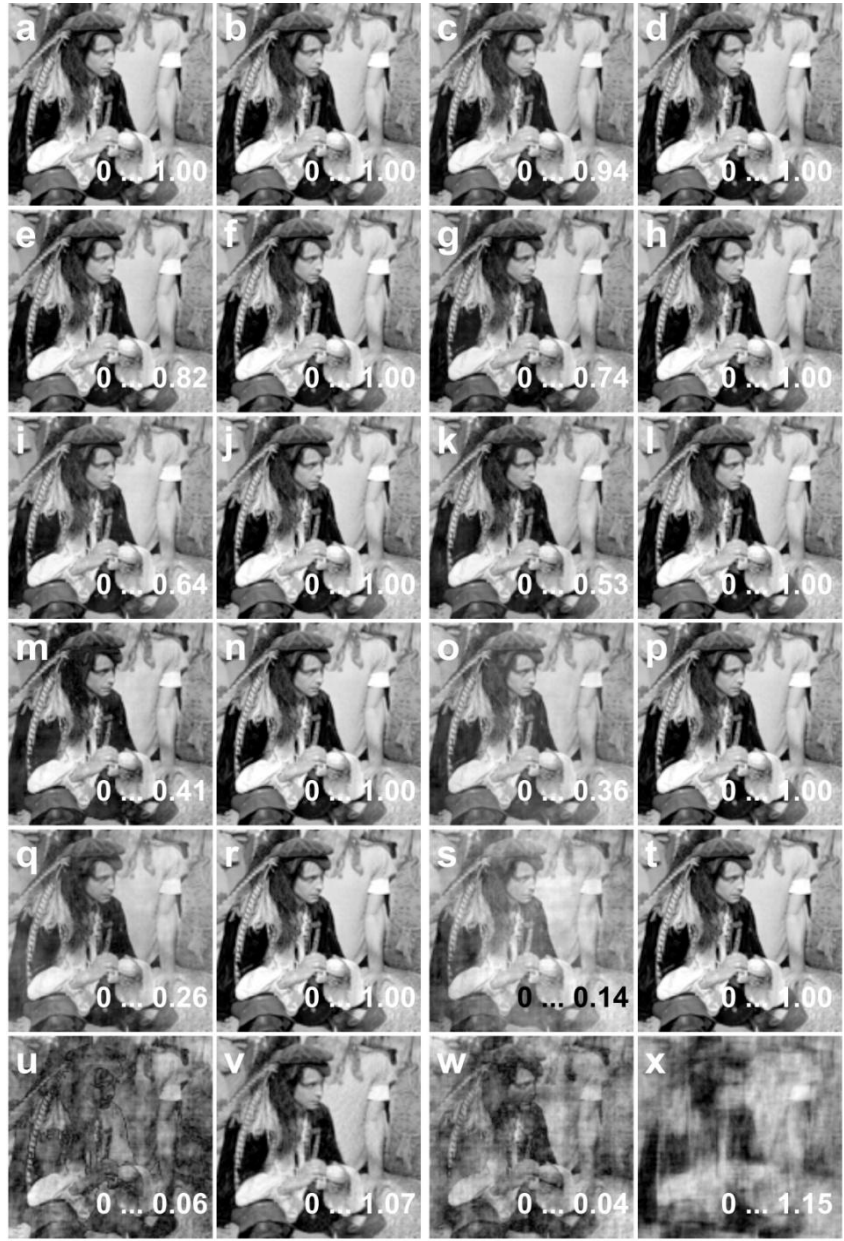

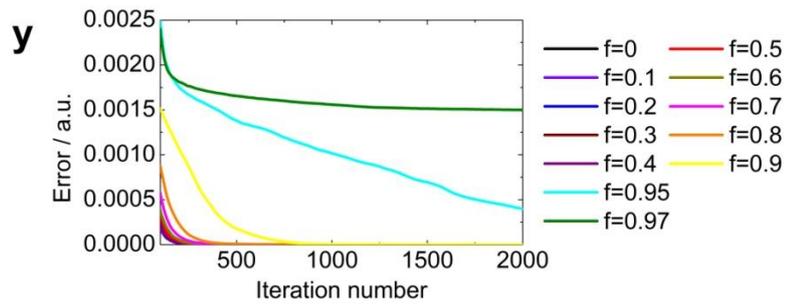

Fig. 5. Missing pixels in coherent diffraction imaging, where diffraction patterns are symmetrised before reconstruction. The central part of the reconstructed object distributions (128 × 128 pixels) is shown. The oversampling ratio is 8, and the total reconstructed object area is sampled with 1024 × 1024 pixels. The reconstructions are shown in pairs. Left: reconstruction obtained from the complex-valued far-field

distribution using the inverse Fourier transform. Right: reconstruction obtained from the far-field diffraction pattern by applying an iterative phase retrieval routine. The numbers in the right bottom corners indicate the amplitude values of the reconstructed object distributions in a.u. The ratio of missing pixels to the total number of pixels ($f$) is (a) – (b) $f=0$, (c) – (d) $f=0.1$, (e) – (f) $f=0.2$, (g) – (h) $f=0.3$, (i) – (j) $f=0.4$, (k) – (l) $f=0.5$, (m) – (n) $f=0.6$, (o) – (p) $f=0.7$, (q) – (r) $f=0.8$, (s) – (t) $f=0.9$, (u) – (v) $f=0.95$, (w) – (x) $f=0.97$. (y) The error as a function of iteration number for different values of $f$, calculated using Eq. 8.

|  | $f=0$ | $f=0.1$ | $f=0.2$ | $f=0.3$ | $f=0.4$ | $f=0.5$ |
|---|---|---|---|---|---|---|
| Inverse FT | 1.46E-11 | 3.43E-4 | 8.64E-4 | 1.28E-3 | 1.54E-3 | 2.02E-3 |
| Iterative | 6.59E-10 | 5.48E-10 | 6.44E-10 | 6.95E-10 | 7.89E-10 | 1.14E-9 |
|  | $f=0.6$ | $f=0.7$ | $f=0.8$ | $f=0.9$ | $f=0.95$ | $f=0.97$ |
| Inverse FT | 2.59E-3 | 2.63E-3 | 3.07E-3 | 3.49E-3 | 3.92E-3 | 3.93E-3 |
| Iterative | 1.15E-9 | 1.94E-9 | 5.79E-9 | 9.24E-8 | 2.47E-4 | 1.05E-3 |

Table 4. Error in the reconstructed object distributions calculated using Eq. 8.

### 2.2.3. "Missing" intensity values in the center of diffraction pattern

In the results shown above, the relation between the number of missing intensity values and the oversampling ratio holds only if the missing pixels are randomly distributed across the diffraction pattern. When the missing intensity values are all located in the centre of the diffraction pattern (as illustrated in Fig. 6), which is often the case in experiment, a meaningful reconstruction cannot be obtained even at small $f$, as shown in Figs. 7 and 8.

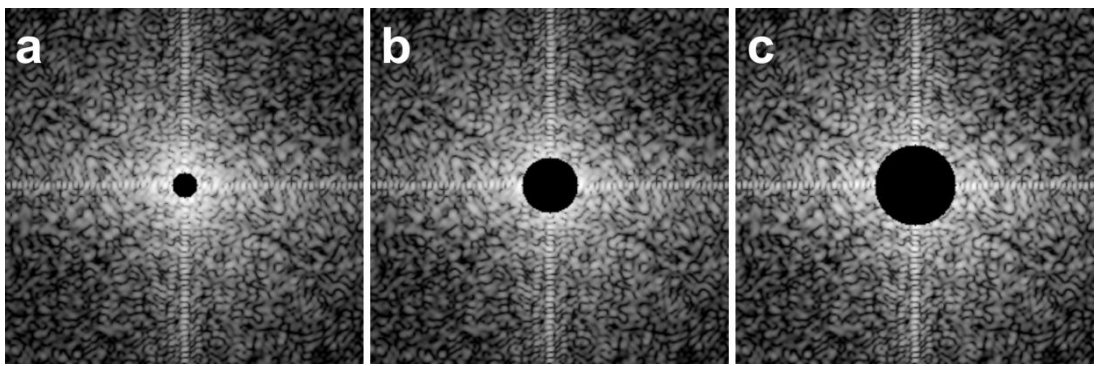

Fig. 6. Missing intensity values in the centre of the diffraction pattern. The object ( "man" image) is sampled with 128 × 128 pixels, the total object area is sampled with 512 × 512 pixels, and the oversampling ratio is 4. Here, only the central 256 × 256 pixels of the simulated diffraction pattern are shown. The ratio of missing pixels is: (a) $f=0.001$, (c) $f=0.005$ and (c) $f=0.01$.

Figures 7 and 8 show reconstructions obtained using inverse FT of complex-valued far-field distributions and the reconstructions obtained by iterative phase retrieval with recovery of the missing intensity values. From the results shown in Figs. 7 and 8 and the errors summarised in Tables 3 and 4, it is apparent that reconstructions of better quality are obtained through iterative phase retrieval than from the inverse FT of complex-valued distributions. It is interesting that despite known phase in the far-field distribution, a single inverse FT delivers a poorer reconstruction than that obtained by iterative phase retrieval of diffraction pattern with missing phases and missing pixels. These results can be explained as follows. The central frequency of the spectrum (diffraction pattern) corresponds to the constant background value in the object distribution. The other frequencies around the central frequency correspond to the low frequencies in the object distribution. When the background and low-frequency information are missing from the object distribution, the resulting object distribution appears in the form of a high-pass filtered original distribution. This effect can be observed in the object reconstructions obtained from an inverse Fourier transform of the complex-valued far-field distributions. When an iterative phase retrieval reconstruction is applied, the missing pixels in the diffraction pattern are recovered, and therefore the background and the low-resolution information in the object distribution are reconstructed.

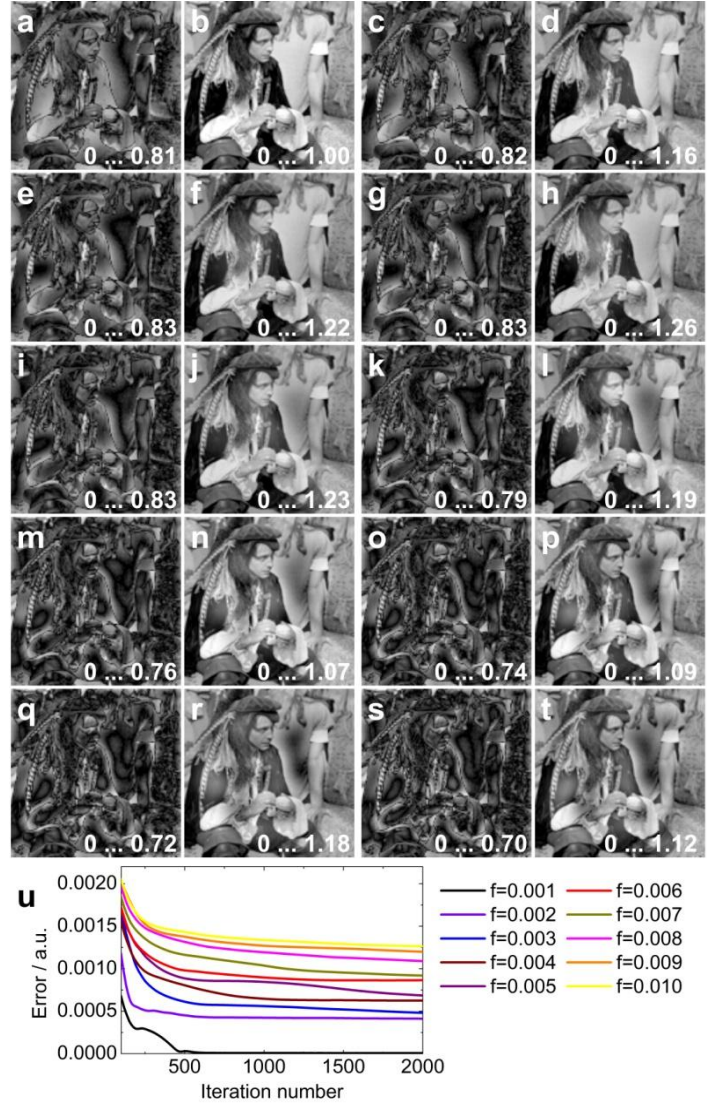

Fig. 7. Missing intensity values in the centre of the diffraction pattern. The central part of the reconstructed distributions (128 × 128 pixels) is shown. The total reconstructed area is sampled with 512 × 512 pixels, and the oversampling ratio is 4. The reconstructions are shown in pairs. Left: reconstruction obtained from the complex-valued far-field distribution by taking the inverse Fourier transform. Right: reconstruction obtained from the far-field diffraction pattern by applying an iterative phase retrieval routine. The numbers in the right bottom corners indicate the amplitude values of the reconstructed object distributions in a.u. The ratio of missing pixels to the total number of pixels ($f$) is (a) – (b) $f=0.001$, (c) – (d) $f=0.002$, (e) – (f) $f=0.003$, (g) – (h) $f=0.004$, (i) – (j) $f=0.005$, (k) – (l) $f=0.006$, (m) – (n) $f=0.007$, (o) – (p) $f=0.008$, (q) – (r) $f=0.009$, (s) – (t) $f=0.01$. (u) The error as a function of iteration number for different values of $f$, calculated using Eq. 8.

|  | $f=0.001$ | $f=0.002$ | $f=0.003$ | $f=0.004$ | $f=0.005$ |
|---|---|---|---|---|---|
| Inverse FT | 2.98E-3 | 3.10E-3 | 3.15E-3 | 3.18E-3 | 3.21E-3 |
| Iterative | 4.95E-6 | 4.09E-4 | 4.78E-4 | 6.24E-4 | 6.81E-4 |
|  | $f=0.006$ | $f=0.007$ | $f=0.008$ | $f=0.009$ | $f=0.010$ |
| Inverse FT | 3.27E-3 | 3.32E-3 | 3.34E-3 | 3.35E-3 | 3.39E-3 |
| Iterative | 8.62E-4 | 9.17E-4 | 1.09E-3 | 1.19E-3 | 1.26E-3 |

Table 5. Error in the reconstructed object distributions calculated using Eq. 8.

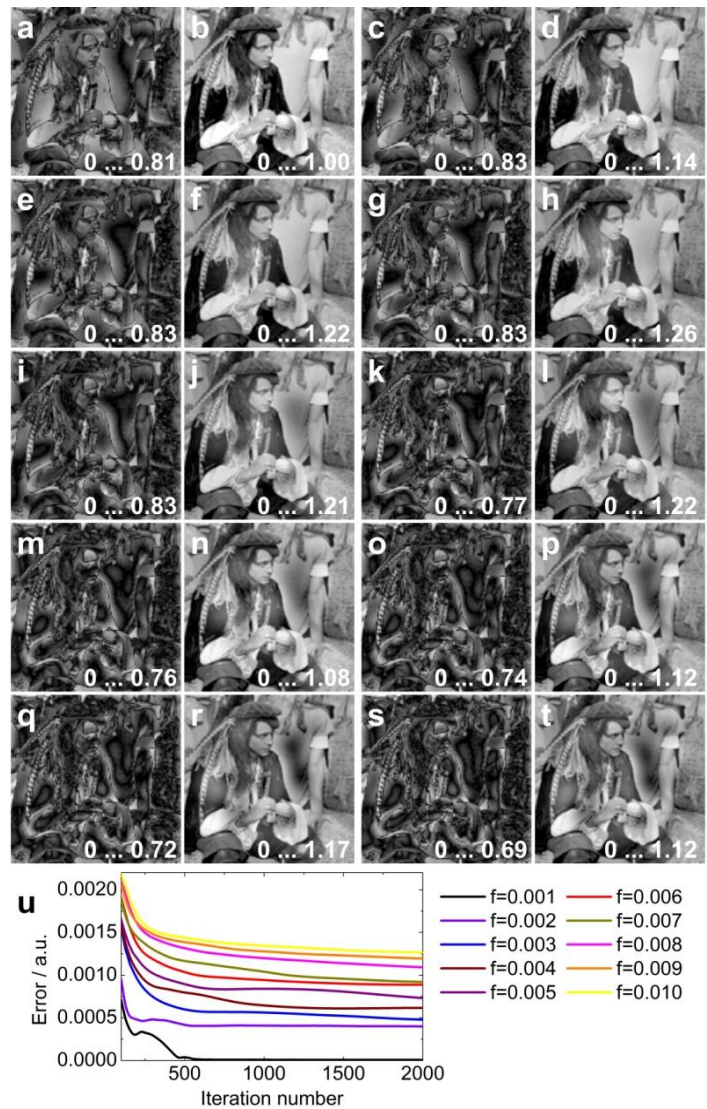

Fig. 8. Missing pixels in the center of the diffraction pattern, where the linear oversampling ratio is 8. The central part of the reconstructed object distributions, 128 × 128 pixels, is shown. The total reconstructed area is sampled with 1024 × 1024 pixels. The reconstructions are shown in pairs. Left: reconstruction obtained from the complex-valued far-field distribution by taking an inverse Fourier transform. Right: reconstruction obtained from the far-field diffraction pattern by applying the iterative phase retrieval routine. The numbers in the right bottom corners indicate the amplitude

values of the reconstructed object distributions in a.u. The ratio of missing pixels to the total number of pixels ($f$) is (a) – (b) $f = 0.001$, (c) – (d) $f = 0.002$, (e) – (f) $f = 0.003$, (g) – (h) $f = 0.004$, (i) – (j) $f = 0.005$, (k) – (l) $f = 0.006$, (m) – (n) $f = 0.007$, (o) – (p) $f = 0.008$, (q) – (r) $f = 0.009$, (s) – (t) $f = 0.01$. (u) The error as a function of the iteration number for different $f$, calculated using Eq. 8.

|            | $f$=0.001 | $f$=0.002 | $f$=0.003 | $f$=0.004 | $f$=0.005 |
|------------|-----------|-----------|-----------|-----------|-----------|
| Inverse FT | 2.98E-3   | 3.10E-3   | 3.15E-3   | 3.19E-3   | 3.22E-3   |
| Iterative  | 5.38E-6   | 3.99E-4   | 4.77E-4   | 6.15E-4   | 7.29E-4   |
|            | $f$=0.006 | $f$=0.007 | $f$=0.008 | $f$=0.009 | $f$=0.010 |
| Inverse FT | 3.29E-3   | 3.32E-3   | 3.34E-3   | 3.36E-3   | 3.39E-3   |
| Iterative  | 8.85E-4   | 9.19E-4   | 1.09E-3   | 1.19E-3   | 1.26E-3   |

Table 6. Error in the reconstructed object distributions calculated using Eq. 8.

From the results shown in Figs. 7 and 8 and Tables 5 and 6, we can conclude that in a situation where the missing intensity values are located in the centre of the diffraction pattern, the magnitude of the linear oversampling ratio does not make a significant difference.

## 2.3. Note on the error metrics

The convergence of the iterative reconstruction process and the quality of the reconstruction is evaluated using an error function. In this study, the original object distribution was available, and the iteratively obtained reconstructed object was therefore compared against the original object distribution by calculating the error as defined by Eq. 8. In reality, the original object distribution is generally unknown, and other error metrics are employed.

An error metric was introduced by Fienup for error-reduction based algorithms in which the error function evaluates how well the iteratively recovered amplitudes match the measured amplitudes in the detector plane [15]:

$$\text{Error} = \left\{ \frac{N^{-2} \sum_{v,w} \left[ |G_k(v,w)| - |F(v,w)| \right]^2}{\sum_{v,w} |F(v,w)|^2} \right\}^{1/2}, \quad (9)$$

where $|F(v,w)|$ are the measured amplitudes, $|G_k(v,w)|$ are the iteratively retrieved amplitudes at the $k$-th iteration, and $(v,w)$ are the coordinates in the detector plane.

Another error metric was introduced by Miao et al [13, 16, 17] for hybrid input-output-based algorithms, in which the error function evaluates how well the recovered object distribution satisfies the object constraints:

$$\text{Error} = \left\{ \frac{\sum_{x,y \notin \gamma} |g'_k(x,y)|^2}{\sum_{x,y \in \gamma} |g'_k(x,y)|^2} \right\}^{1/2}, \quad (10)$$

where $g'_k(x,y)$ is the iteratively reconstructed object distribution at the k-th iteration, $\gamma$ is the support in the object plane, and $(x,y)$ are the coordinates in the object plane.

In the present study, the error was calculated using all three metrics for each reconstruction, as given by Eqs. 8, 9 and 10. It was noted that the reconstructions with the lowest error as defined by Eq. 7 and those with the lowest error as defined by Eqs. 9 and 10 were not the same. For example, for a diffraction pattern with $\sigma = 8$, 10 reconstructions were selected with the lowest error as defined by Eqs. 8, 9 and 10. From the 10 reconstructions evaluated using Eq. 9, only seven were the same as those with the least error evaluated using Eq. 8. From the 10 reconstructions with the lowest error as evaluated using Eq. 10, only three coincided with those with the least error evaluated by Eq. 8. This leads to the conclusion that the error function calculated using Eq. 9 is more precise than the error calculated with Eq. 10.

## 3. "Missing" intensity values in holography

In this section, we investigate the effects of missing intensity values in holography. Although we consider the case of in-line or Gabor-type holography [4, 5], the results obtained here can easily be adapted for other types of holography.

### 3.1. Reference wave extent and "missing" intensity values

An oversampling ratio can be introduced in holography in a similar way as in CDI. It has already been highlighted by Dennis Gabor that in holography, the extent of the reference wave must be larger than the extent of the object wave [5]. This requirement is very similar to the requirement of oversampling in CDI. We therefore introduce the linear oversampling ratio as given by Eq. 3. The measured intensity in the hologram can be written in form of a convolution of the object distribution with the free-space propagator [18]:

$$H(v,w) = |o(v,w) \otimes h(v,w)|^2 \quad (11)$$

which is a set of equations where $o(x,y)$ are the unknowns. For a hologram sampled with $N \times N$ pixels and a complex-valued object sampled with $N_0 \times N_0$ pixels, there are $N^2$ equations and $2N_0^2$ unknowns; for a real-valued object, there are $N^2$ equations and $N_0^2$ unknowns. The system of equations can in principle have a solution if the number of equations exceeds the number of unknowns. For real-valued objects, this condition is:

$$N^2 > N_0^2. \tag{12}$$

When $(1-f)N^2$ pixels are measured, this condition is transformed to:

$$(1-f)N^2 > N_0^2. \tag{13}$$

By substituting the definition of the oversampling ratio provided by Eq. 3 into Eq. 13, we obtain the following condition for the missing pixel ratio in holography:

$$f < 1 - \frac{1}{\sigma^2}. \tag{14}$$

From Eq. 14, for example, we obtain $f < 0.938$ for $\sigma = 4$. This means that for a hologram with a linear oversampling ratio of $\sigma = 4$, the measured intensities for only 6% of all pixels can in principle be sufficient to reconstruct the object distribution and simultaneously recover the intensity at the missing intensity values.

## 3.2. Simulated examples

In-line holograms were simulated assuming the following parameters. A plane wave of wavelength 532 nm propagated through an amplitude object and the resulting hologram was acquired at a distance of 20 mm from the object. The total object area and the hologram both had a size of 2 × 2 mm² and were sampled with 512 × 512 pixels. The object itself was sampled with 128 × 128 pixels, thus giving $\sigma = 4$. The hologram was simulated by applying an angular spectrum method, as explained in Appendix B and in detail elsewhere [18]. The object distribution and the simulated hologram are shown in Figs. 9a and b.

The iterative reconstruction procedure employed here was based on that used for twin image elimination [19], as explained in Appendix B. Holograms with missing intensity values of $f = 0.1...0.98$ were simulated and reconstructed, and not all of them are shown here. A hologram with $f = 0.95$ and its reconstruction are shown in Figs. 9c, d and e. For values of up to $f = 0.9$, the reconstructed object distributions were identical to the original distribution shown in Fig. 9a. At $f = 0.95$, the reconstructed distribution still resembles the original distribution (Fig. 9e), while at $f = 0.98$, the reconstructed distribution does not resemble the original object distribution (Fig. 9f). These results are in good agreement with the theoretical predictions provided by Eq. 14, which indicate that a reconstruction can be obtained if $f < 0.938$.

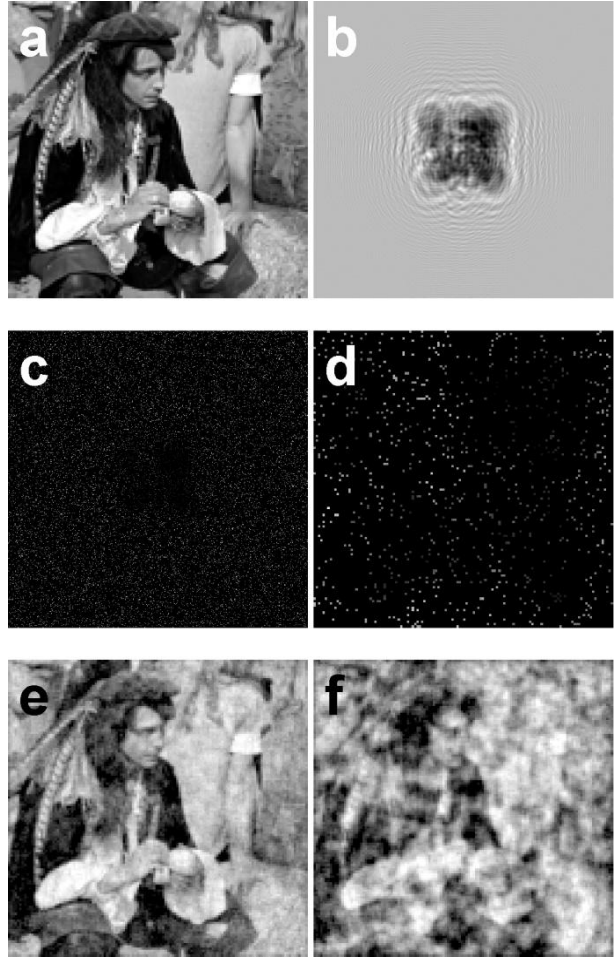

Fig. 9. Missing intensity values in holography. (a) Object distribution sampled with 128 × 128 pixels. (b) Distribution of the simulated hologram, sampled with 512 × 512 pixels. (c) Distribution of the simulated hologram (shown in (b)) with missing intensity values, $f=0.95$. (d) Central part of the hologram shown in (c) (128 × 128 pixels). (e) Reconstruction obtained from the hologram with $f=0.95$ shown in (c). (f) Reconstruction obtained from the hologram with $f=0.98$.

## 4. Relation to compressive sensing

We should point out that the results presented here solve similar problems but are not directly related to the compressive sensing technique [20]. Firstly, in compressive sensing, the governing equations are linear, and the measured and the recovered signals are connected by linear matrix equations. This is not the case for example in CDI, where the phase of the wavefront is missing because only the intensity is measured in the far field, and the object distribution and its diffraction pattern are not connected by a simple linear matrix equation. For holography, under certain approximations and neglecting the conjugated (twin) term, the task of recovering the object distribution from the measured hologram can be written in a form of linear system of equations, meaning that a compressing sensing approach could be applied [21]. Secondly, in compressive sensing, one of the conditions under which recovery is possible is sparsity, which requires the signal to be sparse in some domain. For example, this condition means that the Fourier spectrum of the object consists of a few intense components and that the components at the other frequencies are so small that they can be neglected. This sparsity should not be confused with the missing information. The missing intensities, generally speaking, are not small and cannot be neglected; they are only missing, and for complete recovery of the object distribution, their values (not necessarily negligible) must be recovered. This can be achieved, as shown here, by applying iterative phase retrieval algorithms. However, the sparsity condition may be somewhat related to the oversampling condition, which is the requirement that the object distribution is zero-padded in CDI, and the condition that the extent of the reference wave exceeds the extent of the object in holography. A related problem of reconstructing an object from incomplete frequency samples was addressed by Candes et al [22] who showed that exact recovery may be obtained by solving a convex optimisation problem.

# 5. Conclusions

In this study, we demonstrate that an object can be successfully reconstructed from its diffraction pattern or hologram, even if some intensity values in the diffraction pattern or hologram are missing.

We quantitatively estimate how many of the measured intensity values can be missing. In CDI, for a real-valued object, the ratio of missing pixels to the total number of pixels in the acquired diffraction pattern should not exceed $(1 - 2/\sigma^2)$. Based on this formula, we estimate that even 5% of a noise-free diffraction pattern measured at an oversampling ratio of $\sigma = 8$ is sufficient to simultaneously retrieve the object distribution and the missing intensity values. In holography, for an amplitude object, the ratio of missing pixels to the total number of pixels in the acquired diffraction pattern should not exceed $(1 - 1/\sigma^2)$. Using this formula, we estimate that even 6% of a noise-free hologram measured at an oversampling ratio of $\sigma = 4$ is sufficient to simultaneously retrieve the object distribution and the missing intensity values. We provided the corresponding simulations that confirmed these estimations.

The reconstruction procedure was performed using conventional iterative phase retrieval routines, in which at each iteration, the missing amplitude values were replaced with the iteratively updated values. As a result of this iterative procedure, the object distribution was reconstructed and the missing amplitudes were retrieved. An interesting observation is that even if the phases in the far-field distribution are known, a single inverse FT (a single, non-iterative reconstruction of the hologram) delivers a poorer reconstruction that the that obtained by iterative phase retrieval.

The spatial positions of the missing intensity values in the diffraction pattern play a crucial role in reconstruction. When the missing intensity values are located in the centre of the diffraction pattern, which is often the case in an experiment, a good quality reconstruction cannot be obtained even for a relatively small number of missing pixels.

## Appendix A

The diffraction patterns were simulated and reconstructed as described in detail in [12], here we provide the main details.

**Simulation of the diffraction patterns.** The diffraction patterns were calculated without application of fast Fourier transforms (FFT) to avoid the wrapping of signal at the edge of the images. The simulation procedure was as follows. The object distribution was digitised, that is, it was represented in pixels. For each pixel, the diffracted complex-valued wavefront was calculated as the analytical solution of the diffraction on a square aperture. The total sum of the complex-valued wavefronts from all pixels yielded the total diffracted wavefront. The squared amplitude of the total diffracted wavefront gave the intensity distribution of the diffraction pattern.

**Reconstruction of diffraction patterns.** The simulated diffraction patterns were reconstructed by applying the hybrid input-output (HIO) algorithm with a feedback parameter of 0.9 [15]. One hundred reconstructions were obtained by applying the HIO algorithm with tight object support in the form of a square patch of 128 × 128 pixels under the constraint that the object must be real and positive; a total of 2000 iterations were made. During iterative phase retrieval, the values of the missing pixels in the diffraction pattern were replaced by the values obtained after each iterative loop. Ten reconstructions with the lowest errors (calculated using Eq. 8) were selected, aligned and averaged, thus giving the final reconstruction.

## Appendix B

**Simulation of the in-line holograms.** The in-line holograms were simulated by applying an angular spectrum method (ASM), as explained in more detail elsewhere [18]. Here we provide the main principles. In the ASM [23, 24], the complex-valued exit wave $t(x, y)$ is propagated to the detector plane via the the following transformation [18]:

$$U(X,Y) = \text{FT}^{-1}\left\{\text{FT}[t(x,y)]\exp\left(\frac{2\pi i z}{\lambda}\sqrt{1-\alpha^2-\beta^2}\right)\right\}, \tag{B1}$$

where FT and FT$^{-1}$ are the Fourier transform and inverse Fourier transform, respectively, and $(\alpha, \beta)$ are the Fourier domain coordinates. The Fourier transform is defined as:

$$\text{FT}[t(x,y)] = \iint t(x,y)\exp\left[-2\pi i z\left(\frac{\alpha}{\lambda}x + \frac{\beta}{\lambda}y\right)\right]\text{d}x\text{d}y.$$

The hologram distribution is calculated as

$$H(X,Y) = |U(X,Y)|^2.$$

**Reconstruction of the in-line holograms.** The simulated holograms were reconstructed by applying the iterative phase algorithm [15]. The wavefront was propagated back and forth between the object and the hologram plane. It was propagated from the hologram plane to the object plane by calculating the complex-valued distribution as expressed by Eq. B1, and from the object plane to the hologram plane via the following transformation [18]:

$$t(x,y) = \text{FT}^{-1}\left\{\text{FT}[H(X,Y)]\exp\left(-\frac{2\pi i z}{\lambda}\sqrt{1-\alpha^2-\beta^2}\right)\right\}.$$

Two thousand iterations were applied with tight object support in the form of a square patch of size 128 × 128 pixels under the constraint that the object absorption must be positive. After each 20 iterations, the object distribution was smoothed by calculating the convolution with a smoothing kernel:

$$\text{smooth}[t(x,y)] = \text{FT}^{-1}\left\{\text{FT}[t(x,y)] \cdot \text{FT}\begin{pmatrix} 1 & 1 & 1 \\ 1 & 4 & 1 \\ 1 & 1 & 1 \end{pmatrix}\right\}.$$

During the process of iterative phase retrieval, the values of the missing pixels in the hologram were replaced by the values obtained after each iterative loop.